\documentclass[aps,prl,preprint,groupedaddress]{revtex4}

\usepackage{graphicx}

\usepackage{amsmath,amsthm,amsfonts, latexsym}

\begin{document}


\title{Entanglement and Composite Bosons}


\author{Christopher Chudzicki$^1$, Olufolajimi Oke$^1$, and William K. Wootters$^{1,2}$}
\affiliation{$^1$Department of Physics, Williams College, Williamstown, MA 01267, USA
\\
$^2$Perimeter Institute for Theoretical Physics, 31 Caroline Street North, Waterloo, ON N2L 2Y5, Canada}


\date{\today}

\begin{abstract}
We build upon work by C.~K.~Law [Phys.~Rev.~A {\bf 71}, 034306 (2005)] to 
show in general that the entanglement between two fermions largely determines the extent to which the pair behaves like an elementary boson. Specifically, we derive upper and lower bounds on a quantity
$\chi_{N+1}/\chi_N$ that governs the bosonic character of a pair of fermions when $N$ such pairs approximately share the same wavefunction.  Our bounds depend on 
the purity of the single-particle density matrix, an indicator of entanglement, and demonstrate that if the entanglement is sufficiently strong, the quantity $\chi_{N+1}/\chi_N$ approaches its ideal bosonic value.  
\end{abstract}


\maketitle


Under what circumstances can a pair of fermions be treated as an elementary boson?  Many authors have done detailed studies of this question, as it applies, for example, to atomic Bose-Einstein condensates 
\cite{Avancini, Rombouts}, excitons \cite{Rombouts, Combescotreview, Okumura}, and Cooper pairs in superconductors \cite{Belkhir}.  In a 2005 paper, C.~K.~Law presented evidence that the question can be answered
in general
in terms of {\em entanglement}: two fermions can be treated as an elementary boson if they are sufficiently entangled \cite{Law}.  Consider, for example, a single hydrogen atom in a
harmonic trap.  Within the atom, the proton and electron are strongly entangled with respect to their position variables; for example, wherever the proton might be found---it could be anywhere 
in the trap---the electron is sure to be nearby.  Law suggests that this entanglement is the essential property underlying the
(approximate) bosonic behavior of the composite particle, allowing, for example, a collection of many
hydrogen atoms to form a Bose-Einstein condensate \cite{Kleppner}.  

Specifically, his hypothesis can be expressed as follows: For a single composite particle in a pure state, let $P$ be the {\em purity} of the reduced state 
of either of the two component fermions---$P$ is small when the entanglement between the
two particles is large (see below for the definition)---and let $N$ be the number of composite 
particles that 
approximately share  
the given quantum state.  Then the composite particles can be treated as bosons as 
long as $NP \ll 1$.  That is, according to this hypothesis,
the quantity $1/P$ roughly quantifies the number of particles one can
put into the same pure state, before the composite nature of the particles begins to
interfere appreciably with their ideal bosonic behavior.  

Law's argument in support of this hypothesis assumes a two-particle wavefunction within a certain class, characterized by a specific form of the
eigenvalues of the reduced density matrix of either particle, and he notes that it would be desirable to extend the argument to more general wavefunctions.  Such a generalization is the aim of the present paper.  
With no restrictions on the form of the two-particle wavefunction, we use the purity to place upper and lower 
bounds on Law's measure of bosonic character, and we show that these bounds are the tightest possible of the given form.  In this way we obtain a more general connection between entanglement and
bosonic character.  

Before proceeding to our general argument, it may be instructive to consider the special case
of the hydrogen atom.  Let $\Psi(\vec{R},\vec{r})$ be the wavefunction of a single hydrogen atom in a 
harmonic trap, with $\vec{R}$ and $\vec{r}$ being the position coordinates of the proton and electron, respectively.  For simplicity we assume that the proton is sufficiently massive 
compared to the electron that we
can write this wavefunction as 
\begin{equation}
\Psi(\vec{R},\vec{r}) = \psi(\vec{R})\phi(\vec{r}-\vec{R}),  \label{product}
\end{equation}
where $\psi$ is the ground-state harmonic oscillator wavefunction
\begin{equation}
\psi(\vec{R}) = \frac{1}{\pi^{3/4}b^{3/2}}\exp(-R^2/2 b^2),  \label{harmonic}
\end{equation}
and $\phi$ is the ground-state wavefunction of the electron in a hydrogen atom:
\begin{equation}
\phi(\vec{r}) = \frac{1}{\pi^{1/2} a_0^{3/2}}\exp(-r/a_0).  \label{hydrogen}
\end{equation}
Here $a_0$ is the Bohr radius and $b$ is a length parameter characterizing the size of the trap. 

The purity $P$ of the 
reduced state of either of the two particles is defined by
\begin{equation}
P = \hbox{Tr}\;\rho^2,
\end{equation}
where $\rho$ is the density matrix of the particle in question.  (Because the pair is in a pure state,
the purities of the two particles are guaranteed to be equal.)  Note that $P$ takes values
between zero and one.   For the hydrogen atom, the purity of the proton is given by
\begin{equation}
P = \int \rho(\vec{R}, \vec{R}')\rho(\vec{R}', \vec{R})\, d\vec{R} \, d\vec{R}',
\end{equation}
where the proton's density matrix $\rho$ is
\begin{equation}
\rho(\vec{R},\vec{R}') = \int \Psi(\vec{R},\vec{r})\Psi^*(\vec{R}', \vec{r})\, d\vec{r}.
\end{equation}
Inserting Eq.~(\ref{product}) into the definition of $P$, we find that
\begin{eqnarray}
P &= \int |\psi(\vec{R})|^2 |\psi(\vec{R}')|^2 |\sigma(\vec{R}' - \vec{R})|^2 \, d\vec{R} \, d\vec{R}' 
\label{mainint} \\
&= \int |\psi(\vec{R})|^2 |\psi(\vec{R} + \vec{q})|^2 |\sigma(\vec{q})|^2 \, d\vec{R} \, d\vec{q},
\label{mainint2}
\end{eqnarray}
where
\begin{equation}
\sigma(\vec{q}) = \int \phi(\vec{r})\phi^*(\vec{r} - \vec{q}) \, d\vec{r}.
\end{equation}
But the range of $\sigma$ is comparable to the Bohr radius and much smaller than
the dimension of the trap.  So we can reasonably replace $|\psi(\vec{R}+\vec{q})|^2$
in Eq.~(\ref{mainint2}) with $|\psi(\vec{R})|^2$ and write
\begin{equation}
P = \int |\psi(\vec{R})|^4 \, d\vec{R} \: \int |\sigma(\vec{q})|^2 \, d\vec{q}.
\end{equation}
The integrals can be done, and one finds that 
\begin{equation}
P = \frac{33}{4 \sqrt{2 \pi}} \left(\frac{a_0}{b}\right)^3.
\end{equation}
Thus the purity depends, not surprisingly, on the ratio of the volume of an atom to the volume of the 
trap, and Law's condition $NP \ll 1$ essentially says that the space available to 
each atom must be large compared to its size.  This condition is in rough agreement
with the condition that the number of atoms be small compared to the maximum 
occupation number as computed in Ref.~\cite{Rombouts}.


We now turn to the general argument.

Consider a composite particle formed from two distinguishable, fundamental fermions $A$ and 
$B$ with wavefunction $\Psi(x_A, x_B)$.  (Here the $x$'s could be vectors in any number
of dimensions.)  Writing this wavefunction in its Schmidt decomposition
yields
\begin{equation}
\Psi(x_A, x_B) = \sum_p \lambda_p^{1/2} \phi_p^{(A)}(x_A) \phi_p^{(B)}(x_B) .
\end{equation}
Here $\phi_p^{(A)}$ and $\phi_p^{(B)}$ are the Schmidt modes, constituting orthonormal bases for the 
states of particles $A$ and $B$, and the $\lambda_p$'s, which are the eigenvalues
of each of the single-particle density matrices, are nonnegative real numbers satisfying 
$\sum_p \lambda_p = 1$.  In terms of the $\lambda_p$'s, the purity can be written as 
\begin{equation}
P = \sum_p \lambda_p^2 .
\end{equation}
Again, a small value of the purity indicates a large entanglement.

In terms of creation operators, the state $\Psi(x_A, x_B)$ can be written as
\begin{equation}
\Psi(x_A, x_B) = \sum_p \lambda_p^{1/2} a_p^\dag b_p^\dag |0\rangle ,
\end{equation}
where $a_p^\dag$ creates an $A$ particle in the state $\phi_p^{(A)}(x_A)$, $b_p^\dag$ creates a $B$
particle in the state $\phi_p^{(B)}(x_B)$, and $|0\rangle$ is the vacuum state.
The composite particle creation operator $c^\dag$, which creates a pair of $A$ and $B$
particles in the state $\Psi(x_A, x_B)$, is defined to be
\begin{equation}
c^\dag = \sum_p \lambda_p^{1/2} a_p^\dag b_p^\dag .
\end{equation}
Our analysis, like Law's, aims to determine to what extent the operators $c^\dag$ and $c$ act like bosonic creation and annihilation operators when applied to a state consisting of $N$ composite particles.  

Consider the state obtained by antisymmetrizing the product state
$\Psi(x_{A}^{(1)}, x_{B}^{(1)}) \cdots \Psi(x_{A}^{(N)}, x_{B}^{(N)})$.  In terms of the creation operator $c^\dag$, 
we can write the properly antisymmetrized state as 
\begin{equation}
|N\rangle = \frac{1}{\sqrt{N!}} \chi_N^{-1/2} (c^\dag)^N |0\rangle.  \label{Ndef}
\end{equation}
Here $\chi_N$ is a normalization constant necessary because $c^\dag$ is not a perfect
bosonic creation operator.  The quantity $\chi_N$ is given by \cite{Combescot, Law}
\begin{equation}
\chi_N = \frac{1}{N!} \langle 0 | c^N (c^\dag)^N |0\rangle = \sum_{\begin{array}{c} \vspace{-11mm} \\ \hbox{\scriptsize $p_1 \ldots p_N$}\\ \vspace{-11mm} \\ \hbox{\scriptsize all different} \end{array}}
\lambda_{p_1} \lambda_{p_2} \ldots \lambda_{p_N} .
\end{equation}
(This expression gives $\chi_N = 0$ if the number $N$ exceeds the number of Schmidt
modes with nonzero Schmidt coefficient.  In that case $(c^\dag)^N |0\rangle = 0$ and we cannot define
the state $|N\rangle$.)
For an ideal boson, we would have $\chi_N = 1$.  

Note that $c^\dag|N\rangle$ is not necessarily equal to
$\sqrt{N+1}\,|N+1\rangle$.
Rather, it follows from the definition (\ref{Ndef}) that
\begin{equation}
c^\dag|N\rangle = \alpha_{N+1} \sqrt{N+1}\,|N+1\rangle,
\end{equation}
where
\begin{equation}
\alpha_{N} = \sqrt{\frac{\chi_{N}}{\chi_{N-1}}}\, .  \label{1}
\end{equation}
Similarly, instead of $c|N\rangle = \sqrt{N}|N-1\rangle$, we have
\begin{equation}
c|N\rangle = \alpha_N \sqrt{N} |N-1\rangle + |\epsilon_N\rangle ,
\end{equation}
where $|\epsilon_N\rangle$ is orthogonal to $|N-1\rangle$.  For perfect bosons, we would have $\langle \epsilon_N |
\epsilon_N \rangle = 0$, but the actual value is  \cite{Combescot, Law}
\begin{equation}
\langle \epsilon_N | \epsilon_N \rangle = 1 - \frac{\chi_{N+1}}{\chi_N}
 - N \left( \frac{\chi_{N}}{\chi_{N-1}} 
- \frac{\chi_{N+1}}{\chi_N} \right).   \label{2}
\end{equation}

In the Appendix, we show that the ratio $\chi_{N+1}/\chi_N$ which appears in 
Eqs.~(\ref{1}) and (\ref{2}) is strictly non-increasing as $N$ increases (more precisely, we show
that $\chi_N^2 - \chi_{N+1}\chi_{N-1}$ is non-negative), so that the quantity in parentheses
in Eq.~(\ref{2}) is non-negative.  It follows that 
both $\alpha_N$ and $\langle \epsilon_N|\epsilon_N\rangle$ will be within a small amount $\delta$ of their bosonic values 
when  $\chi_{N+1}/\chi_N \ge 1-\delta$.  One can also show \cite{Combescot, Law}
that
\begin{equation}
\langle N| [c,c^\dag] |N\rangle = 2\left( \frac{\chi_{N+1}}{\chi_N}  \right) -1 ,
\end{equation}
which is within $2\delta$ of its ideal bosonic value, 1,  under the same condition.
We therefore follow Law in using the ratio
$\chi_{N+1}/\chi_N$---we call it the ``$\chi_N$-ratio''---as our indicator of bosonic
character \cite{Law,Law2}.  

One might wonder why we confine our attention to quantities involving only the state $|N\rangle$
and nearby states, rather than insisting that the operator $c$ act like a bosonic 
operator on the whole 
subspace spanned by $\{|0\rangle, \ldots, |N\rangle \}$.  The reason is that we are interested
in a state that approximates $|N\rangle$, and we wish to quantify the degree to which 
the system behaves like a collection of bosons when a composite particle is added to
or removed from this state.  Hence our focus on $\chi_{N+1}/\chi_N$ as 
the quantifier of bosonic character rather than $\chi_N$ itself.  
We note that because the $\chi_N$-ratio is non-increasing with $N$, a lower 
bound on $\chi_{N+1}/\chi_N$ will also be a lower bound on $\chi_{N'+1}/\chi_{N'}$ for all 
$N' < N$.  However, as one can see in Ref.~\cite{Combescot}, this fact is not sufficient to guarantee that $\chi_N$ itself is close to unity 
whenever $\chi_{N+1}/\chi_N$ is.

In the remainder of the paper we prove two inequalities relating
the $\chi_N$-ratio to the purity.  

The first is a lower bound: $\chi_{N+1}/\chi_N \ge 1-NP$.
To show this, we consider the quantity $\chi_{N+1} - \chi_N(1 - NP)$ and show that it
must be non-negative.
\begin{equation} 
\begin{split}
\chi_{N+1} - \chi_N(1 - NP) = 
\sum_{\begin{array}{c} \vspace{-11mm} \\ \hbox{\scriptsize $p_1 \ldots p_{N+1}$}\\ \vspace{-11mm} \\ \hbox{\scriptsize all different} \end{array}}
\lambda_{p_1} \lambda_{p_2} \ldots \lambda_{p_{N+1}}
- \left( 1 - N \sum_p \lambda_p^2 \right) 
\sum_{\begin{array}{c} \vspace{-11mm} \\ \hbox{\scriptsize $p_1 \ldots p_N$}\\ \vspace{-11mm} \\ \hbox{\scriptsize all different} \end{array}}
\lambda_{p_1} \lambda_{p_2} \ldots \lambda_{p_N}  \\
= 
\sum_{\begin{array}{c} \vspace{-11mm} \\ \hbox{\scriptsize $p_1 \ldots p_{N+1}$}\\ \vspace{-11mm} \\ \hbox{\scriptsize all different} \end{array}}
\lambda_{p_1} \lambda_{p_2} \ldots \lambda_{p_{N+1}}
- \sum_{\begin{array}{c} \vspace{-11mm} \\ \hbox{\scriptsize $p_1 \ldots p_N$}\\ \vspace{-11mm} \\ \hbox{\scriptsize all different;} \\ \vspace{-11mm} \\ \hbox{\scriptsize $p_{N+1}$ free} \end{array}}
\lambda_{p_1} \lambda_{p_2} \ldots \lambda_{p_N}\lambda_{p_{N+1}}
+ N
\sum_{\begin{array}{c} \vspace{-11mm} \\ \hbox{\scriptsize $p_1 \ldots p_N$}\\ \vspace{-11mm} \\ \hbox{\scriptsize all different;} \\ \vspace{-11mm} \\ \hbox{\scriptsize $p_{N+1}$ free} \end{array}}
\lambda_{p_1} \lambda_{p_2} \ldots \lambda_{p_N}\lambda_{p_{N+1}}^2
\end{split}
\end{equation}
Note that the first two sums of the last line
have many terms in common, which therefore cancel out.  The only
terms remaining from those sums are the terms in the second sum for which the 
value of $p_{N+1}$ is equal to the value of one of the indices $p_k$ with $k=1, \ldots , N$.  Each of these
$N$ possibilities yields the same result; so we can combine those first two sums into
the expression
\begin{equation}
-N \sum_{\begin{array}{c} \vspace{-11mm} \\ \hbox{\scriptsize $p_1 \ldots p_{N}$}\\ \vspace{-11mm} \\ \hbox{\scriptsize all different} \end{array}} \lambda_{p_1}^2 \lambda_{p_2} \cdots \lambda_{p_N}.
\end{equation}
We therefore have
\begin{equation}
\chi_{N+1} - \chi_N(1 - NP) = 
-N \sum_{\begin{array}{c} \vspace{-11mm} \\ \hbox{\scriptsize $p_1 \ldots p_{N}$}\\ \vspace{-11mm} \\ \hbox{\scriptsize all different;}\\  \vspace{-11mm} \\ \hbox{\scriptsize $p_{N+1}$ free} \end{array}} \lambda_{p_1}^2 \lambda_{p_2} \cdots \lambda_{p_N}
\lambda_{p_{N+1}}
+ N
\sum_{\begin{array}{c} \vspace{-11mm} \\ \hbox{\scriptsize $p_1 \ldots p_N$}\\ \vspace{-11mm} \\ \hbox{\scriptsize all different;} \\ \vspace{-11mm} \\ \hbox{\scriptsize $p_{N+1}$ free} \end{array}}
\lambda_{p_1} \lambda_{p_2} \ldots \lambda_{p_N}\lambda_{p_{N+1}}^2.
\end{equation}
Again the two sums have many terms in common.  Cancelling these terms leaves
\begin{equation}  \label{xxx}
\begin{split}
\chi_{N+1} - \chi_N(1 - NP) =
N(N-1) 
\sum_{\begin{array}{c} \vspace{-11mm} \\ \hbox{\scriptsize $p_1 \ldots p_N$}\\ \vspace{-11mm} \\ \hbox{\scriptsize all different} \end{array}}
\lambda_{p_1}^3 \lambda_{p_2} \ldots \lambda_{p_N} 
- N(N-1) \sum_{\begin{array}{c} \vspace{-11mm} \\ \hbox{\scriptsize $p_1 \ldots p_N$}\\ \vspace{-11mm} \\ \hbox{\scriptsize all different} \end{array}}
\lambda_{p_1}^2 \lambda_{p_2}^2 \ldots \lambda_{p_N}   \\
=
N(N-1) \sum_{\begin{array}{c} \vspace{-11mm} \\ \hbox{\scriptsize $p_1 \ldots p_N$}\\ \vspace{-11mm} \\ \hbox{\scriptsize all different} \end{array}}
\lambda_{p_1} \lambda_{p_2} \ldots \lambda_{p_N} 
\left( \lambda_{p_1}^2 -  \lambda_{p_1} \lambda_{p_2} \right).
\end{split}
\end{equation}
Now, Eq.~(\ref{xxx}) can be rewritten as
\begin{equation}  \label{twenty-one}
\begin{split}
\chi_{N+1} - \chi_N(1 - NP) = 
\frac{N(N-1)}{2}
\sum_{\begin{array}{c} \vspace{-11mm} \\ \hbox{\scriptsize $p_1 \ldots p_N$}\\ \vspace{-11mm} \\ \hbox{\scriptsize all different} \end{array}}
\lambda_{p_1} \lambda_{p_2} \ldots \lambda_{p_N}
\left(\lambda_{p_1}^2 +  \lambda_{p_2}^2
- 2 \lambda_{p_1}\lambda_{p_2} \right)    \\
= \frac{N(N-1)}{2} \sum_{\begin{array}{c} \vspace{-11mm} \\ \hbox{\scriptsize $p_1 \ldots p_N$}\\ \vspace{-11mm} \\ \hbox{\scriptsize all different} \end{array}}
\lambda_{p_1} \lambda_{p_2} \ldots \lambda_{p_N}
\left(\lambda_{p_1} - \lambda_{p_2} \right)^2
\ge 0 ,
\end{split}
\end{equation}
thus yielding the bound 
\begin{equation} \label{first}
\frac{\chi_{N+1}}{\chi_{N}} \ge 1 - NP.
\end{equation}

This bound shows that a sufficiently small purity entails nearly bosonic character as
quantified by $\chi_{N+1}/\chi_{N}$.  We now derive a bound in the other direction,
showing that
a nearly bosonic value of $\chi_{N+1}/\chi_{N}$ implies a small purity.  For this purpose we
start with 
\begin{equation} 
\begin{split}
(1-P)\chi_N - \chi_{N+1} = \left( 1 - \sum_p \lambda_p^2 \right)
\sum_{\begin{array}{c} \vspace{-11mm} \\ \hbox{\scriptsize $p_1 \ldots p_N$}\\ \vspace{-11mm} \\ \hbox{\scriptsize all different} \end{array}}
\lambda_{p_1} \lambda_{p_2} \ldots \lambda_{p_N} -
\sum_{\begin{array}{c} \vspace{-11mm} \\ \hbox{\scriptsize $p_1 \ldots p_{N+1}$}\\ \vspace{-11mm} \\ \hbox{\scriptsize all different} \end{array}}
\lambda_{p_1} \lambda_{p_2} \ldots \lambda_{p_{N+1}} \\
= \sum_{\begin{array}{c} \vspace{-11mm} \\ \hbox{\scriptsize $p_1 \ldots p_N$}\\ \vspace{-11mm} \\ \hbox{\scriptsize all different;} \\ \vspace{-11mm} \\ \hbox{\scriptsize $p_{N+1}$ free} \end{array}}
\lambda_{p_1} \lambda_{p_2} \ldots \lambda_{p_N}\lambda_{p_{N+1}}
- \sum_{\begin{array}{c} \vspace{-11mm} \\ \hbox{\scriptsize $p_1 \ldots p_N$}\\ \vspace{-11mm} \\ \hbox{\scriptsize all different;} \\ \vspace{-11mm} \\ \hbox{\scriptsize $p_{N+1}$ free} \end{array}}
\lambda_{p_1} \lambda_{p_2} \ldots \lambda_{p_N}\lambda_{p_{N+1}}^2
- \sum_{\begin{array}{c} \vspace{-11mm} \\ \hbox{\scriptsize $p_1 \ldots p_{N+1}$}\\ \vspace{-11mm} \\ \hbox{\scriptsize all different} \end{array}}
\lambda_{p_1} \lambda_{p_2} \ldots \lambda_{p_{N+1}} 
\end{split}
\end{equation}
By combining sums as before (inserting the identity $1 = \sum_{p_{N+1}} 
\lambda_{p_{N+1}}$ when needed), we 
get
\begin{equation}   \label{fourteen}
(1-P)\chi_N - \chi_{N+1} = (N-1) \sum_{\begin{array}{c} \vspace{-11mm} \\ \hbox{\scriptsize $p_1 \ldots p_{N+1}$}\\ \vspace{-11mm} \\ \hbox{\scriptsize all different} \end{array}}
\lambda_{p_1}^2 \lambda_{p_2} \ldots \lambda_{p_{N+1}}
+ N(N-1) \sum_{\begin{array}{c} \vspace{-11mm} \\ \hbox{\scriptsize $p_1 \ldots p_N$}\\ \vspace{-11mm} \\ \hbox{\scriptsize all different} \end{array}}
\lambda_{p_1}^2 \lambda_{p_2}^2 \ldots \lambda_{p_N} \ge 0 .
\end{equation}
Combining this result with our earlier inequality (Eq.~(\ref{first})), we have
\begin{equation}  \label{twenty-two}
1-NP \le \frac{\chi_{N+1}}{\chi_N} \le 1 - P.
\end{equation}

We have thus put upper and lower bounds on the $\chi_N$-ratio of a composite particle made of 
two distinguishable fermions, in terms of the entanglement of the pair.  We have not specified 
anything about the form of the wavefunction of the composite particle; so the link between the $\chi_N$-ratio
and entanglement is established in general.


The lower bound in Eq.~(\ref{twenty-two}) is in fact as strong a bound as one could hope to derive in terms of purity, in that the bound is achievable: if there are $M$ nonzero Schmidt
modes and $\lambda_p = 1/M$, then, by Eq.~(\ref{twenty-one}), $\chi_{N+1}/\chi_N = 1 - NP$
as long as $N$ is less than $M$.
This lower bound is also achieved by wavefunctions in the class Law considers---this class includes double Gaussian wavefunctions---in the limit $NP \ll 1$.  Because Eq.~(\ref{fourteen}) is never zero unless $N=1$ (in which case it is
always zero), our upper bound is not, for general $N$, achievable.  Nevertheless, it is the best
possible
upper bound of the form $\chi_{N+1}/\chi_N \le 1 - bP$, whether or not $b$ depends on $N$.  This is because for any value of $b$ greater than 1, there exists a distribution of Schmidt coefficients that 
makes $1-bP$ negative---it suffices to make one of the coefficients $\lambda_k$ very large---whereas $\chi_{N+1}/\chi_N$ is certainly non-negative.  We note also that there can be no upper bound of the form $1-bP^r$ with $r$ less than 1, because such a bound would contradict our lower bound when $P$ is small.  

We have considered in this paper only a single wavefunction $\Psi(x_A, x_B)$ of the composite particle.
One would also like to investigate whether, for several orthogonal wavefunctions $\Psi_j(x_A, x_B)$, the corresponding creation
operators $c_j^\dag$ approximately satisfy the bosonic relation  $[c_j, c_k^\dag] = 0$ for
$j\ne k$.  (The relation $[c_j, c_k]=0$ will automatically be satisfied because
of the anticommutation of the underlying fermionic operators.)
If the relevant deviation from this commutation relation similarly diminishes to zero as the entanglement of each wavefunction increases, one will then have further evidence for 
the proposition that entanglement is crucial
for determining whether a pair of fermions can be treated as a boson.  

Taking this idea to its logical conclusion, Law notes that two particles can be highly entangled even if they are far apart.  Could we treat such a pair of fermions as a composite boson?  The above analysis suggests that we can do so.  However, we would have to regard the pair as a very {\em fragile} boson in the absence of an interaction that would preserve the pair's entanglement in the face of external disturbances.  On this view, the role of interaction in creating a composite boson is not fundamentally to keep the two particles close to each other, but to keep them entangled.

\begin{acknowledgments}
We thank Fred Strauch, as well as the participants of the quantum information group at Perimeter Institute, for their stimulating questions and valuable suggestions.  Research
at Perimeter Institute is supported by the Government of Canada through Industry Canada and by the Province of Ontario through the Ministry of Research and Innovation.  
\end{acknowledgments}

\section{Appendix: Proof that $\chi_{N}^2 - \chi_{N+1}\chi_{N-1} \ge 0$}
Let us use the symbol $\sum'$ to indicate a sum over all the indices appearing in the summand, with the restriction that they must all take distinct values.  We have, then,
\begin{eqnarray}
\chi_N^2 - \chi_{N+1}\chi_{N-1} = \hbox{$\sum'$} \lambda_{r_1}\cdots \lambda_{r_N}
\hbox{$\;\sum'$} \lambda_{p_1}\cdots \lambda_{p_N}
-\hbox{$\;\sum'$} \lambda_{r_1}\cdots \lambda_{r_{N+1}}
\hbox{$\;\sum'$} \lambda_{p_1}\cdots \lambda_{p_{N-1}}.
\end{eqnarray}
We now treat separately the sum over $p_{N}$ in the first term and the sum over
$r_{N+1}$ in the second term, obtaining
\begin{eqnarray}
\chi_N^2 - &\chi_{N+1}\chi_{N-1} = \hbox{$\;\sum'$}  \lambda_{r_1}\cdots \lambda_{r_N}
\left[ \hbox{$\;\sum'$}  \lambda_{p_1}\cdots \lambda_{p_{N-1}}
- \hbox{$\;\sum'$} \lambda_{p_1}\cdots \lambda_{p_{N-1}}\left(\lambda_{p_1}+
\cdots + \lambda_{p_{N-1}}\right) \right] \nonumber \\
\hfill &- \left[\hbox{$\;\sum'$}  \lambda_{r_1}\cdots \lambda_{r_{N}}
-\hbox{$\;\sum'$}  \lambda_{r_1}\cdots \lambda_{r_{N}}\left(\lambda_{r_1}+
\cdots + \lambda_{r_{N}}\right) \right]
\hbox{$\;\sum'$}  \lambda_{p_1}\cdots \lambda_{p_{N-1}} \hfill  \label{yyyy} \\
& = \hbox{$\;\sum'$}  \lambda_{r_1}\cdots \lambda_{r_N} 
\hbox{$\;\sum'$}  \lambda_{p_1}\cdots \lambda_{p_{N-1}}
\left[\lambda_{r_1} + \cdots + \lambda_{r_{N}} - \left( \lambda_{p_1}
+ \cdots + \lambda_{p_{N-1}}\right)\right]  \nonumber.
\end{eqnarray}
We now separate the sum over the $p$'s into two parts: (i) the part in which
$p_1, \ldots, p_{N-1}$ have the same restrictions as $r_1, \ldots, r_{N-1}$,
and (ii) the part in which one of the $p$'s has the same value as $r_{N}$.
So the expression becomes
\begin{eqnarray}
\hbox{$\sum$} \lambda_{r_N}\;\hbox{$\sum'_{\ne r_N}$}\lambda_{r_1}\cdots\lambda_{r_{N-1}}
\;\hbox{$\sum'_{\ne r_N}$}\lambda_{p_1}\cdots\lambda_{p_{N-1}}
\left[ \lambda_{r_N} + \left( \lambda_{r_1} + \cdots + \lambda_{r_{N-1}}\right)
- \left( \lambda_{p_1} + \cdots + \lambda_{p_{N-1}}\right)\right] \nonumber  \\
+ (N-1) \hbox{$\sum$} \lambda_{r_N}^2\;\hbox{$\sum'_{\ne r_N}$}\lambda_{r_1}\cdots\lambda_{r_{N-1}}
\;\hbox{$\sum'_{\ne r_N}$}\lambda_{p_1}\cdots\lambda_{p_{N-2}}
\left[ \left( \lambda_{r_1} + \cdots + \lambda_{r_{N-1}}\right)
- \left( \lambda_{p_1} + \cdots + \lambda_{p_{N-2}}\right)\right]. \nonumber
\end{eqnarray}
In the first line, everything in the square bracket vanishes except for
$\lambda_{r_N}$, because of the symmetry between the $r$'s and the $p$'s.
Thus the first line is non-negative.  Meanwhile the expression in the second line that appears
within the
sum over $r_N$ has the same form as the last line of Eq.~(\ref{yyyy}), except
with one fewer $r$ index and one fewer $p$ index.  Therefore the same maneuver
can be repeated again and again, until at last we are left with an expression
that is manifestly non-negative, because there are no more of the negative $p$ terms.  It follows that
\begin{equation}
\chi_N^2 - \chi_{N+1}\chi_{N-1} \ge 0.
\end{equation}


\end{document}